\documentclass[journal,twoside]{IEEEtran}
\usepackage{amsmath}
\usepackage{algorithmic}
\usepackage{array}
\usepackage{url}
\usepackage{color}
\usepackage[switch,columnwise]{lineno}
\usepackage{mathtools}
\usepackage[noadjust]{cite}
\usepackage[numbers,sort&compress,square]{natbib}
\newcommand{\etal}{\mbox{\emph{et al.\ }}}
\usepackage{amssymb}
\usepackage{todonotes}
\usepackage{graphicx}
\graphicspath{{images/}{../images/}}
\usepackage{subfigure}
\usepackage{xcolor}
\usepackage[export]{adjustbox}
\usepackage{relsize}
\usepackage{soul,xcolor}
\usepackage{hyperref}
\usepackage{algorithm,algorithmic}
\usepackage{calc}  
\usepackage{enumitem} 

\DeclareMathOperator*{\argmin}{arg\,min}

\newcommand{\estimage}{x^\textnormal{(image)}}
\newcommand{\estdata}{x^\textnormal{(data)}}
\newcommand{\cevar}{x}
\newcommand{\proxvar}{v}
\newcommand{\sensoragent}{\textnormal{s}}
\newcommand{\imageagent}{\textnormal{i}}
\newcommand{\dataagent}{\textnormal{d}}
\newcommand{\Amat}{\mathbf{A}}
\newcommand{\measdata}{y}
\newcommand{\stackedcevar}{\underline{\mathbf{\cevar}}}

\begin{document}


\title{Data and Image Prior Integration for Image Reconstruction Using Consensus Equilibrium}
\author{Muhammad Usman Ghani, \IEEEmembership{Student Member,~IEEE,}
        W. Clem Karl,~\IEEEmembership{Fellow,~IEEE}
        \thanks{This material is based upon work supported by the U.S. Department of Homeland Security, Science and Technology Directorate, Office of University Programs, under Grant Award 2013-ST-061-ED0001. The views and conclusions contained in this document are those of the authors and should not be interpreted as necessarily representing the official policies, either expressed or implied, of the U.S. Department of Homeland Security.}
        
\thanks{M. U. Ghani and W. C. Karl are with the Department
of Electrical and Computer Engineering, Boston University, Boston,
MA, 02215 USA (e-mail: \texttt{\{mughani, wckarl\}@bu.edu)}.}
}

\markboth{Submitted to IEEE Transactions on Computational Imaging}%
{Ghani and Karl: Data and Image Prior Integration for Image Reconstruction Using Consensus Equilibrium}

%



\maketitle

\begin{abstract}
Image domain prior models have been shown to improve the quality of reconstructed images, especially when data are limited. Pre-processing of raw data, through the implicit or explicit inclusion of data domain priors have separately also shown utility in improving reconstructions. In this work, a principled approach is presented allowing the unified integration of \emph{both} data and image domain priors for improved image reconstruction. The consensus equilibrium framework is extended to integrate physical sensor models, data models, and image models. In order to achieve this integration, the conventional image variables used in consensus equilibrium are augmented with variables representing data domain quantities. The overall result produces combined estimates of both the data and the reconstructed image that is consistent with the physical models and prior models being utilized. The prior models used in both domains in this work are created using deep neural networks. The superior quality allowed by incorporating both data and image domain prior models is demonstrated for two applications: limited-angle CT and accelerated MRI. The prior data model in both these applications is focused on recovering missing data. Experimental results are presented for a $90^0$ limited-angle tomography problem from a real checked-bagged CT dataset and a $4\times$ accelerated MRI problem on a simulated dataset. The new framework is very flexible and can be easily applied to other computational imaging problems with imperfect data. 

\end{abstract}

\IEEEpeerreviewmaketitle

\section{Introduction}
\ifCLASSOPTIONcaptionsoff
  \newpage
\fi

\IEEEPARstart{C}{onventional}, analytical image formation algorithms, such as the filtered-back projection algorithm (FBP) assume that high quality data is present on a dense and regular grid. In certain situations, however it is either impossible or undesirable to fulfill these data acquisition requirements. Examples include low-dose computed tomography (CT) \cite{chang2017modeling,ghani2018cnn}, sparse-view CT \cite{chun2018convolutional,ghani2018deep,ye2018deepGlobalsip}, limited-angle CT \cite{anirudh2017lose, wurfl2018deep, huang2019data, bubba2019learning}, accelerated MRI \cite{han2019k, jacob2019structured, haldar2013low}, diverging-wave ultrasound \cite{lu2020reconstruction,ghani2019high}, single-pixel imaging \cite{duarte2008single}, Fourier Ptychography \cite{tian2014multiplexed} and interrupted synthetic aperture radar (SAR) \cite{cetin2014sparsity}. Using conventional image reconstruction methods with such imperfect data produces images filled with artifacts that are difficulty to interpret. 

Model-based image reconstruction (MBIR) methods provide an alternative approach to conventional, analytical image formation methods which explicitly incorporate physical sensor and image prior models. Image prior models capture desirable image features, which enable MBIR methods to produce higher quality image reconstructions \cite{jin2015model, elbakri2002statistical, de2000reduction}. A variety of image priors, including Total-Variation (TV) \cite{ritschl2011improved}, Markov Random Field (MRF) models \cite{zhang2016gaussian}, and  deep-learning-based prior models \cite{ye2018deep} have been explored. Even simple prior models such as TV have been shown to greatly improve image quality, though at the expense of significantly increased computation. An alternative approach has been to focus on transforming the given observed data to better meet the assumptions underlying fast conventional analytical reconstruction methods. In particular, these approaches pre-process the physical observations with the goal of producing data estimates of high quality on dense and regular grids. This technique has been done by using data-domain prior models in estimation frameworks focused entirely in the data domain \cite{la2005penalized, jacob2019structured, haldar2013low, ghani2018deep}. Such data-domain approaches have been shown to be computationally efficient and capable of yielding high-quality resulting imagery, though often of lesser quality than image-domain MBIR methods.\par 

If incorporating prior models in the data domain improves image quality and prior models in the image domain also improves image quality, a natural question is whether better overall images could be obtained by incorporating \emph{both} types of prior models in a unified framework. Preliminary work involving partial inclusion of both types of information has suggested there may indeed be benefit of such integration \cite{ghani2019integrating}. In this work, we present a principled method to integrate both data and image domain prior models in an image reconstruction framework. Intuitively, combining both types of prior models will allow us to incorporate more prior knowledge and therefore should result in better reconstructions. The proposed framework uses consensus equilibrium (CE) \cite{buzzard2018plug} to combine models of sensor, data, and image to obtain a unified reconstructed image combining all sources of information. Consensus Equilibrium is itself a framework to combine multiple heterogeneous agents in generating an image estimate. To exploit the machinery of CE here, the original image variable is augmented with a data variable so that each CE agent updates estimates of both image \emph{and} data. Inspired by the maximum a posteriori (MAP) estimation theory, three CE agents are proposed: the first agent is based on the physical sensor model, the second agent is based on an image prior model and the third agent is based a data prior model. In this work limited-angle CT and accelerated MRI are used as a prototype problems, however, our proposed framework is very general and can be easily applied to other computational imaging problems with imperfect data.

\subsection{Contributions}

The major contributions of this work include:

\subsubsection{A Unified Framework for Integration of Data and Image Priors} The main contribution of this work is a general and flexible framework that integrates both data-domain and image-domain priors for image reconstruction in a balanced and principled way. 

\subsubsection{Specification and Use of Rich Data and Image Domain Priors Based on Deep Learning} State of the art conditional generative adversarial network (cGAN) based deep learning models are used for generation of both data-domain and image-domain prior models. These state of the art models capture a rich range of data and image properties. 

\subsubsection{Comparison of Explicit and Implicit Data-domain Priors}
Two different strategies to incorporate data-domain models are examined and compared. The first approach is based on the use of the proximal map associated to an explicitly defined regularized data estimation problem. The alternative approach uses a deep neural network (DNN) to directly perform data denoising and enhancement, thus incorporating an implicit data prior. 

\subsubsection{Comparison of Framework Effectiveness on Canonical Imperfect Data Applications} We demonstrate the effectiveness of our proposed framework on two canonical imperfect data applications: i) $90^0$ limited angle CT , and ii) $4\times$ accelerated MRI. We also provide comparison to a number of popular alternative approaches. Our framework outperforms existing image post-processing and state-of-the-art MBIR approaches and methods using either data or image priors alone. This demonstrates the unifying and general nature of our proposed framework.

\section{Related Work}
An overview of recent advances in model-based imaging and data-domain models is presented in this section. Using implicitly or explicitly defined image priors in a model-based image reconstruction (MBIR) framework has been a popular theme in  recent years. The plug-and-play framework (PnP-MBIR) \cite{venkatakrishnan2013plug} uses ADMM to split the original problem into sensor and image-domain model sub-problems. It does not require the image priors to be explicitly defined, therefore an off-the-shelf image denoiser can be used instead of solving an expensive prior-regularized image sub-problem. Similar strategies have been used with other formulations for variable splitting and replacement of image prior proximal maps by learned models \cite{ye2018deep,meinhardt2017learning,ono2017primal,kamilov2017plug,gupta2018cnn}. The RED method \cite{romano2017little,reehorst2018regularization} adopts a similar strategy except that it explicitly defines the image-domain regularizer. These approaches provide principled methods with high resulting imaging quality by coupling a physically accurate imaging model and a powerful image prior. \par
   
Data-domain models and processing methods have also been proposed which couple some form of raw data enhancement with conventional analytical image reconstruction algorithms. For example, \cite{ghani2018deep} used a trained DNN to complete missing projection data and then used the filtered back projection (FBP) algorithm for image reconstruction of sparse-view CT. Structured low-rank matrix-based methods have been used in various MRI applications to perform k-space data completion or correction \cite{haldar2013low,jacob2019structured}. Han \etal \cite{han2019k} have used data-domain deep learning for k-space data completion. Once k-space data is completed or corrected, these approaches use an inverse Fourier transform for image reconstruction. Jin \etal \cite{jin2016compressive} used a structured low-rank matrix-based approach to complete randomly sub-sampled ultrasound data measurements and then applied delay-and-sum (DAS) beamforming for image reconstruction.
   
Our initial work in \cite{ghani2019integrating} explored the potential of combining both data and image domain models to produce higher quality images than PnP-MBIR alone \cite{venkatakrishnan2013plug, ye2018deep}. That preliminary work demonstrated the potential of combining models in both domains, though the data-domain component was effectively limited to a pre-processing step. In this work we extend that aim by providing a principled and integrated approach to incorporating both data and image domain models on an equal footing into the image reconstruction process through extensions of the consensus equilibrium approach. 

\subsection{Consensus Equilibrium Overview}

The recently proposed consensus equilibrium (CE) method \cite{buzzard2018plug} is used in this work to create a framework that integrates both data and image priors in a single unified way. Consensus equilibrium generalizes the plug-and-play framework \cite{sreehari2016plug}, extends it beyond optimization, and allows the integration of multiple sources of information captured through “agents" or mappings. It defines a set of equilibrium conditions which lead to a consensus solution for all considered sources of information. Given a set of $N$ vector valued agents or maps $F_i(x_i)$ of images $x_i$, the consensus equilibrium image $x^*$ of the agents is defined as a solution of the set of equations:
\begin{eqnarray}
	F_{i}(x_i^*) & = & x^*, \; i=1,\ldots,N \\
	\sum_{i=1}^N \mu_i x_i^* & = & x^*
\end{eqnarray}
where $\mu_i$ defines the relative contribution of each agent $F_i$ to the overall solution with $\sum_{i=1}^N \mu_i = 1$. Further details of the CE method can be found in \cite{buzzard2018plug}. In current applications of CE the variables $\cevar_i$ and $\cevar$ are taken to be image domain variables and the agents are chosen as proximal operators associated to data likelihoods or image regularization operators or perhaps just image denoisers. In particular, the variables and mappings are restricted to image domain mappings. In this work, we extend the approach to include data-domain mappings.

\section{Combining Data and Image Priors through Consensus Equilibrium} \label{dip_mbir}

The method proposed here uses the CE approach to integrate both data-domain and image-domain priors.  In order to achieve this aim, an image-domain variable $\estimage$ is augmented with a data-domain variable $\estdata$:
\begin{equation}
    \cevar = \left( 
    \begin{array}{c}
    \estimage\\
	\estdata
	\end{array}
    \right)
\end{equation}
so the unknown CE variable $\cevar$ now contains information about both the data and image domains. If the length of $\estimage$ is $N_i$ and the length of $\estdata$ is $N_d$, then the length of the overall CE estimation variable $\cevar$ is $N=N_i+N_d$. We will denote the image and data components of such augmented variables with superscript (image) and (data) labels, respectively.

Three CE agents are now defined to incorporate information about the problem under consideration. The first agent $F_\sensoragent$ focuses on capturing information about the physical sensing process. The second agent $F_\imageagent$ focuses on prior information about the underlying image. The third agent $F_\dataagent$ incorporates prior information in the data domain about the sensor data. The corresponding CE equations defining the consensus solution $\cevar^*$ for these agents are then given by:
\begin{equation}
	\label{eqn:cesoln}
    \begin{aligned}
    	 F_\sensoragent (\cevar_\sensoragent^*) & = & \cevar^*  \\
		F_\imageagent(\cevar_\imageagent^*)  & = & \cevar^*  \\
		F_\dataagent(\cevar_\dataagent^*)  & = & \cevar^*  \\
		\mu_\sensoragent \cevar_\sensoragent^*+ \mu_\imageagent \cevar_\imageagent^* + \mu_\dataagent \cevar_\dataagent^* & = & \cevar^*     
    \end{aligned}
\end{equation}
where $\cevar_\sensoragent$, $\cevar_\imageagent$, and $\cevar_\dataagent$ are auxiliary variables associated to each agent. Note that because of augmentation all the ``$x$" variables in these equations represent both image and data domain components of a solution and all are of length $N$. 

\begin{figure}[tb]
    \centering
    \includegraphics[width=0.4\textwidth]{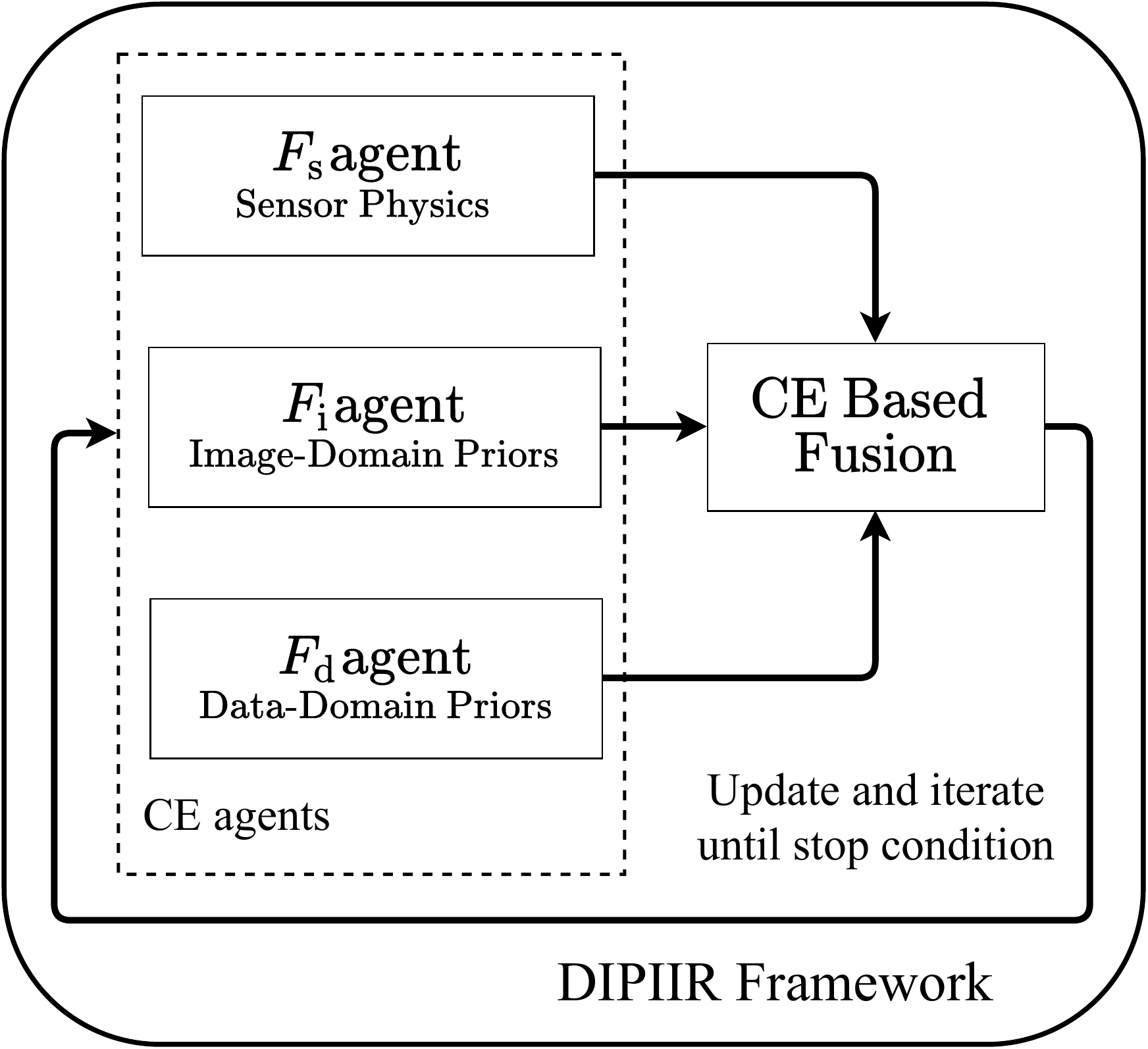}\\
    \caption{Overview of our DIPIIR framework is presented. Data and image domain priors are integrated into Model-based image reconstruction (MBIR) using consensus equilibrium framework.}
    \label{fig:dip_mbir}
\end{figure}

Figure \ref{fig:dip_mbir} presents an overview of our new framework, which we term \emph{Data and Image Prior Integration for Image Reconstruction} (DIPIIR).  The sensor agent incorporates the physical sensing model and imposes consistency with observed data on the estimates. In other words, it improves image and data estimates by pulling initial estimates towards the sensor manifold. The prior agents, on the other hand, impose structural or feature consistency on the resulting estimates based on information we encode about the behavior of images and corresponding data. Intuitively, these prior models project the estimate onto a ``prior manifold". Overall, all three agents combine the sensor physics, image prior, and data prior models to enhance the estimated data and image quality. Finally, the CE equations guide the solution towards consensus of all three agents. Next, we describe our initial choices of these agents in more detail. \par

\subsection{Sensor-based Agent} \label{sensor_agent_mbir}

The sensor-based agent is chosen as a proximal map arising from the data-fidelity term seen in MAP-type estimation:
\begin{equation}
	\label{eq:model_agent}
	\begin{aligned}
	F_\sensoragent(\cevar_\sensoragent) = 
	& \argmin_{\proxvar \geqslant 0} \| \measdata - \Amat \proxvar \|_W^2 + \lambda_\sensoragent \|\proxvar - \cevar_\sensoragent\|_2^2\\
	\end{aligned}
\end{equation}
where the vector $\measdata\in \mathbb{R}^{M}$ is related to the measured data in an application appropriate way, the operator $\Amat \in \mathbb{R}^{(M\times N)}$ incorporates information about the physical sensing operator as well as constraints relating image to data, $\lambda_\sensoragent$ is a trade-off parameter, and $W \in \mathbb{R}^{(M \times M)}$ is a diagonal data weighting matrix allowing weighting of differing data reliability. Note that the optimization variable $\proxvar$ in (\ref{eq:model_agent}) itself is an augmented variable and carries information about both data and image domain quantities. In Section~\ref{sec:incompletedatamodel} we provide details of specific choices for $\measdata$ and $\Amat$ for incomplete-data applications demonstrating how such sensor-related image and data constraints can be flexibly included in the proposed framework. While (\ref{eq:model_agent}) corresponds to a Gaussian noise model, other types of log-likelihood terms are possible (e.g. Poisson). 

\subsection{Data-domain Prior Agent}

A key novelty of this work is the introduction of a data-domain prior agent. This agent incorporates important information about the data variable which helps to progressively improve the data estimates. Two strategies for this agent are explored: an explicit and an implicit data-domain prior.  

\subsubsection{Explicit Data-Domain Prior}
This strategy defines the data agent through a familiar explicit, MAP-like, cost minimization as follows:
\begin{equation}
	\label{eq:data-explicit}
	F_\dataagent(\cevar_\dataagent)
	=
	\argmin_{\proxvar \geqslant 0}  \| \proxvar_0^{\textnormal{(data)}} - \mathbf{S} \proxvar \|_2^2 
	+ 
	\lambda_\dataagent \| \proxvar - \cevar_\dataagent \|_2^2
\end{equation}
where $\mathbf{S}= [\mathbf{0}_{N_i}, \mathbf{I}_{N_d}]$ is a selection operator that extracts the data domain component from the augmented state and $\proxvar_0^{\textnormal{(data)}}$ is a static prior on the data component of the optimization variable $\proxvar$. The idea is that the prior  $\proxvar_0^{\textnormal{(data)}}$ be chosen as an enhanced version of the original observed data. For incomplete data problems, $\proxvar_0^{\textnormal{(data)}}$ could be the output of a data completion deep network, similar to \cite{ghani2019integrating}. For highly noisy or blurry data problems, $\proxvar_0^{\textnormal{(data)}}$ can be the output of a data enhancement deep network or even a simple filtering operation. In both scenarios, the data enhancement operation creating $\proxvar_0^{\textnormal{(data)}}$ is performed once and $\proxvar_0^{\textnormal{(data)}}$ is static throughout the iterative optimization process of CE. Note that overall the agent $F_\dataagent(\cevar_\dataagent)$ only operates on the data domain variable $\estdata_\dataagent$ and simply copies the current estimate of the image variable $\estimage_\dataagent$ to its output. The minimization in (\ref{eq:data-explicit}) can be solved in closed form yielding the following expression for the action of this explicit sensor agent: 
\begin{equation}
	\begin{aligned}
		F_\dataagent(\cevar_\dataagent)	
		=
		\left[
			\begin{array}{c}
			\estimage_\dataagent \\ \hline
			\frac{
					\proxvar_0^{\textnormal{(data)}} + \lambda_\dataagent \cevar^{\textnormal{(data)}}_\dataagent
					}{
					1 +\lambda_\dataagent 
				} 
			\end{array}
		\right]
	\end{aligned}
\end{equation}

\subsubsection{Implicit Data-Domain Prior}
Rather than defining the action of the agent through solution of a minimization problem, in this approach we directly define a mapping through a data-enhancement DNN creating an implicit prior. In particular, the action of this agent is given by:
\begin{equation}
	\label{eq:data-implicit}
	F_\dataagent(\cevar_\dataagent)
	=
		\left[
			\begin{array}{c}
			\estimage_\dataagent \\ \hline
			\psi_{\dataagent}(\estdata_\dataagent)
			\end{array}
		\right]	
\end{equation}
where $\psi_{\dataagent}$ is a data enhancement DNN created from training data to improve partial and degraded observations. As in the explicit case, this agent is crafted to only operate on the estimated data variable $\estdata_\dataagent$ and simply copies the current estimate of image variable $\estimage_\dataagent$ to its output.

Before proceeding we note that other forms of data-domain prior information could be incorporated in the data-domain agent, such as structured-low rank models \cite{jacob2019structured,haldar2013low}. 

\subsection{Image-domain Prior Agent}

Our framework could accommodate a variety of prior agents for the image domain agent. In the current work we have chosen to use a DNN that is trained to perform image enhancement and applied it to the image-domain component of the augmented state $\psi_\imageagent(\estimage_\imageagent)$. Thus the image-domain prior agent $F_\imageagent$ only operates on this image variable $\estimage_\imageagent$ and simply copies the current estimate of data variable $\estdata_\imageagent$. Overall, the action of this agent is:
\begin{equation}
	F_\imageagent(\cevar_\imageagent)
	=
	\left[
		\begin{array}{c}
		\psi_\imageagent(\estimage_\imageagent)
		\\ \hline
		\estdata_\imageagent
		\end{array}
	\right]	
\end{equation}
Such an implicit, DNN-derived image prior offers flexibility and the ability to capture rich image behaviors  \cite{ye2018deep,venkatakrishnan2013plug}. 

Note that it would be straightforward to define the image agent as a proximal map associated with an image-domain regularization process (c.f. MAP estimation): 
\begin{equation}\label{eq:image_agent_generic}
\begin{aligned}
	F_\imageagent(\cevar_\imageagent) = 
	\argmin_{\proxvar \geqslant 0} \phi_\imageagent(\proxvar) + \lambda_\imageagent \| \proxvar - \cevar_\imageagent \|_2^2
\end{aligned}
\end{equation}
where $\lambda_\imageagent$ is the trade-off parameter and $\phi_\imageagent$ is chosen as a regularization penalty, for example derived from methods such as total-variation (TV) \cite{ritschl2011improved}, or Markov random fields models \cite{zhang2016gaussian}.

\subsection{DIPIIR Algorithm}

The solution of the CE equations are provided in \cite{buzzard2018plug}, which we summarize and apply here. First, define the stacked set of consensus agent auxiliary variables:
\begin{equation}
	\stackedcevar
	=
	\left[
	\begin{array}{c}
		\cevar_\sensoragent \\
		\cevar_\dataagent \\
		\cevar_\imageagent
	\end{array}
	\right]
\end{equation}
Recall that each of the individual elements of this vector are augmented to have both an image and a data component. Now define a corresponding vectorized agent map $\mathbf{F}$:
\begin{equation}
	\mathbf{F}(\stackedcevar) =  
	\Bigg(
	\begin{tabular}{c}
		$F_\sensoragent(\cevar_\sensoragent)$\\
		$F_\dataagent(\cevar_\dataagent)$\\
		$F_\imageagent(\cevar_\imageagent)$\\
	\end{tabular}
	\Bigg)
\end{equation}
Finally define the following weighted averaging and redistribution operator $\mathbf{G}$:
\begin{equation}
	\mathbf{G}(\stackedcevar) =   
	\Bigg(
	\begin{array}{c}
    \langle \stackedcevar \rangle\\
	\langle \stackedcevar \rangle\\
	\langle \stackedcevar \rangle
	\end{array}
	\Bigg)
\end{equation}
where $\langle \stackedcevar \rangle = \mu_\sensoragent \cevar_\sensoragent + \mu_\dataagent \cevar_\dataagent + \mu_\imageagent \cevar_\imageagent$ is a weighted average of the components in $\stackedcevar$. 

A solution of the CE equations (\ref{eqn:cesoln}) can be obtained by finding a fixed point $\stackedcevar^*$ of the map $\mathbf{T} = (2\mathbf{F}-\mathbf{I})(2\mathbf{G}-\mathbf{I})$. Once $\stackedcevar^*$ is found, a CE solution $\cevar^*$ can be easily computed from the fixed point as a weighted average of its components: $\cevar^* = \langle \stackedcevar^* \rangle =  \mu_\sensoragent \cevar_\sensoragent^* + \mu_\dataagent \cevar_\dataagent^* + \mu_\imageagent \cevar_\imageagent^*$. The image and data estimates are then just sub-components of this vector. 

One way to achieve this fixed point $\stackedcevar^*$ is through Mann iterations \cite{buzzard2018plug}:
\begin{equation}
	\stackedcevar^{(k+1)} = (1-\rho) \stackedcevar^{(k)} +  \rho \mathbf{T} \stackedcevar^{(k)}
\end{equation}
for all $k \geq 0$, and $\rho \in (0,1)$, where $\stackedcevar^{(0)}$ is an initial estimate. This approach leads to the Algorithm~1.  
\begin{algorithm}
	\caption{DIPIIR Algorithm for Image Reconstruction}
	\begin{algorithmic}[1]
		\label{algo}
		\renewcommand{\algorithmicrequire}{\textbf{Input:}}
		\renewcommand{\algorithmicensure}{\textbf{Output:}}
		\REQUIRE $\measdata$, $\lambda_\sensoragent$, $\lambda_\dataagent$ (If needed)
		\ENSURE  ${\estimage}^*$ (reconstruction), ${\estdata}^*$ (estimated data)
		
		\STATE \textit{CE Initialization:} \\
		$\stackedcevar^{(0)} \xleftarrow{} \textnormal{a value} \in \mathbb{R}^{3N} $\\
		$k \xleftarrow{} 0$\\
		\STATE \textit{CE Solution:} \\
		\WHILE{not converged}
		\STATE $\underline{\mathbf{v}} \xleftarrow{} \left(2 \mathbf{G}(\stackedcevar^{(k)}) - \stackedcevar^{(k)} \right) $
		\STATE $\underline{\mathbf{z}} \xleftarrow{} (2 \mathbf{F}(\underline{\mathbf{v}}) - \underline{\mathbf{v}} ) $ 
		\STATE $\stackedcevar^{(k+1)} \xleftarrow{} (1-\rho) \stackedcevar^{(k)} + \rho \underline{\mathbf{z}}$
		\STATE $k \xleftarrow{} k + 1$
		\ENDWHILE
		\RETURN $\cevar^* \xleftarrow{} \langle \stackedcevar^k \rangle$
	\end{algorithmic} 
\end{algorithm}

\begin{figure*}
	\centering
	\includegraphics[width=0.9\textwidth]{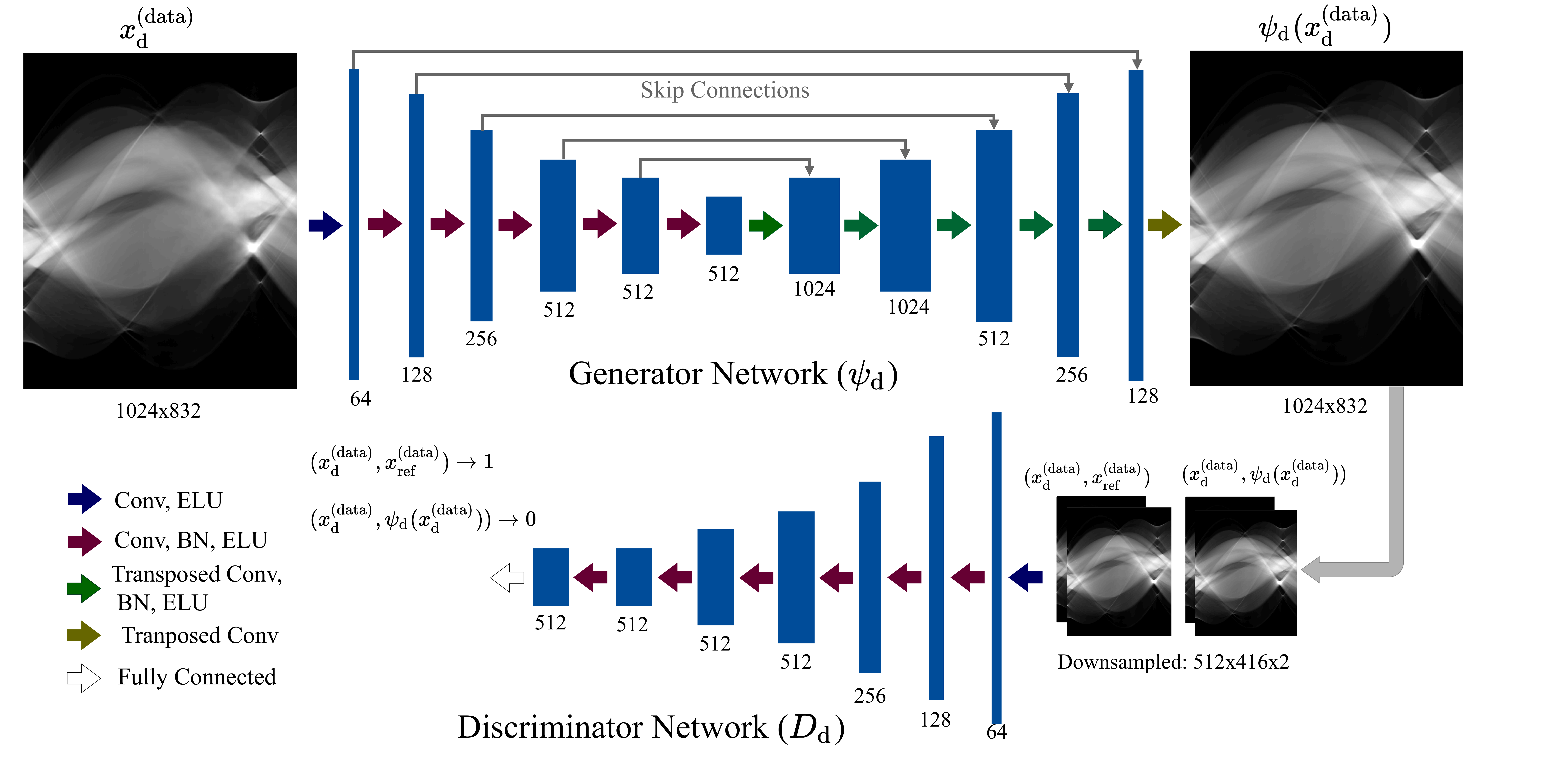}
	\caption{Overall architecture of the data enhancement cGAN $\psi_{\dataagent}(\estdata_\dataagent)$. Here, $\estdata_\textnormal{ref}$ is reference high-quality sensor training data. The abbreviated legends in the Figure are defined here; Conv: 2D convolution, ELU: exponential linear unit, BN: batch-normalization, and 2D Transposed Conv: transposed convolution.}
	\label{fig:cgan_unet}
\end{figure*}

\section{Sensor-based Agent for Problems with Incomplete Data} 
\label{sec:incompletedatamodel}

In this section examples of how the sensor agent $F_\sensoragent(\cevar_\sensoragent)$ can be crafted for problems with incomplete data are provided, and in particular, choices for $\measdata$ and $\Amat$ are given. Applications with incomplete data are an important class of problems and would include sparse-view CT \cite{chun2018convolutional,ghani2018deep,ye2018deepGlobalsip}, limited-angle CT \cite{anirudh2017lose, wurfl2018deep}, accelerated MRI \cite{han2019k, jacob2019structured, haldar2013low}, diverging-wave ultrasound \cite{lu2020reconstruction,ghani2019high}, interrupted SAR \cite{cetin2014sparsity}, and image inpainting \cite{bertalmio2000image} to name a few. The experimental results we provide in Section~\ref{sec:exp}, focus on incomplete data problems in CT and MRI. 

One way to cast such problems in the proposed framework is to let the data domain variable $\estdata$ represent the unobserved or missing part of the data and then define the sensing vector $\measdata$ and sensing operator $\Amat$ as follows:
\begin{eqnarray}
	\measdata 
	& = &  
	\left[
	\begin{array}{c}
	y_\text{obs} \\ \hline
	\underline{0}
	\end{array}
	\right] \\
	\Amat 
	&=&  
	\left[ 
	{\begin{array}{cc}
	\mathbf{A}_\text{obs}   & \mathbf{0}_p \\
	\mathbf{A}_\text{unobs} & -\mathbf{I}_p \\
	\end{array} } 
	\right]
\end{eqnarray}
where $y_\text{obs}$ is the physically observed data, $\mathbf{A}_\text{obs}$ captures the physical map from the underlying image $\estimage$ to the observed data, and $\mathbf{A}_\text{unobs}$ reflects the mapping of the image to the unobserved part of the data domain. The unobserved data can be missing projections in the case of limited-data CT, missing Fourier samples for accelerated MRI, or missing pixel values for image inpainting. To better understand the effect of these choices we incorporate them into (\ref{eq:model_agent}) and rewrite the resulting sensor agent. Assuming $W=I$ for simplicity, we obtain: 
\begin{align}
	\label{eq:missingdata_agent}
	&
	F_\sensoragent(\cevar_\sensoragent) = \\ \nonumber
	&
	\argmin_{\proxvar \geqslant 0} 
	\| y_\text{obs} - \mathbf{A}_\text{obs} \proxvar^\text{(image)} \|^2 
	+
	\| \proxvar^\text{(data)} - \mathbf{A}_\text{unobs} \proxvar^\text{(image)} \|^2 
	\\ \nonumber
	& ~~~~~~~~~~
	+ 
	\lambda_\sensoragent \|\proxvar - \cevar_\sensoragent\|_2^2
\end{align}
The first term in (\ref{eq:missingdata_agent}) couples the observed data to the underlying image estimate through the physical sensing model. The second term couples the image estimate and the missing data estimate through the prediction provided by the sensing operator. In particular, as the estimate of the missing data improves it should contribute to the estimate of the underlying image. Note that since (\ref{eq:missingdata_agent}) is quadratic, it can be solved in closed form. In practice, however, when dealing with large imaging problems iterative methods such as conjugate gradient are used. 

\section{Learned Data and Image Models} \label{DL_models}

Our framework refers to two learned DNN models, $\psi_{\dataagent}(\estdata_\dataagent)$ and $\psi_\imageagent(\estimage_\imageagent)$, used to define data and image priors respectively. The details of these prior DNN models are described in the following sections. 

\begin{figure*}
	\centering
	\includegraphics[width=0.7\textwidth]{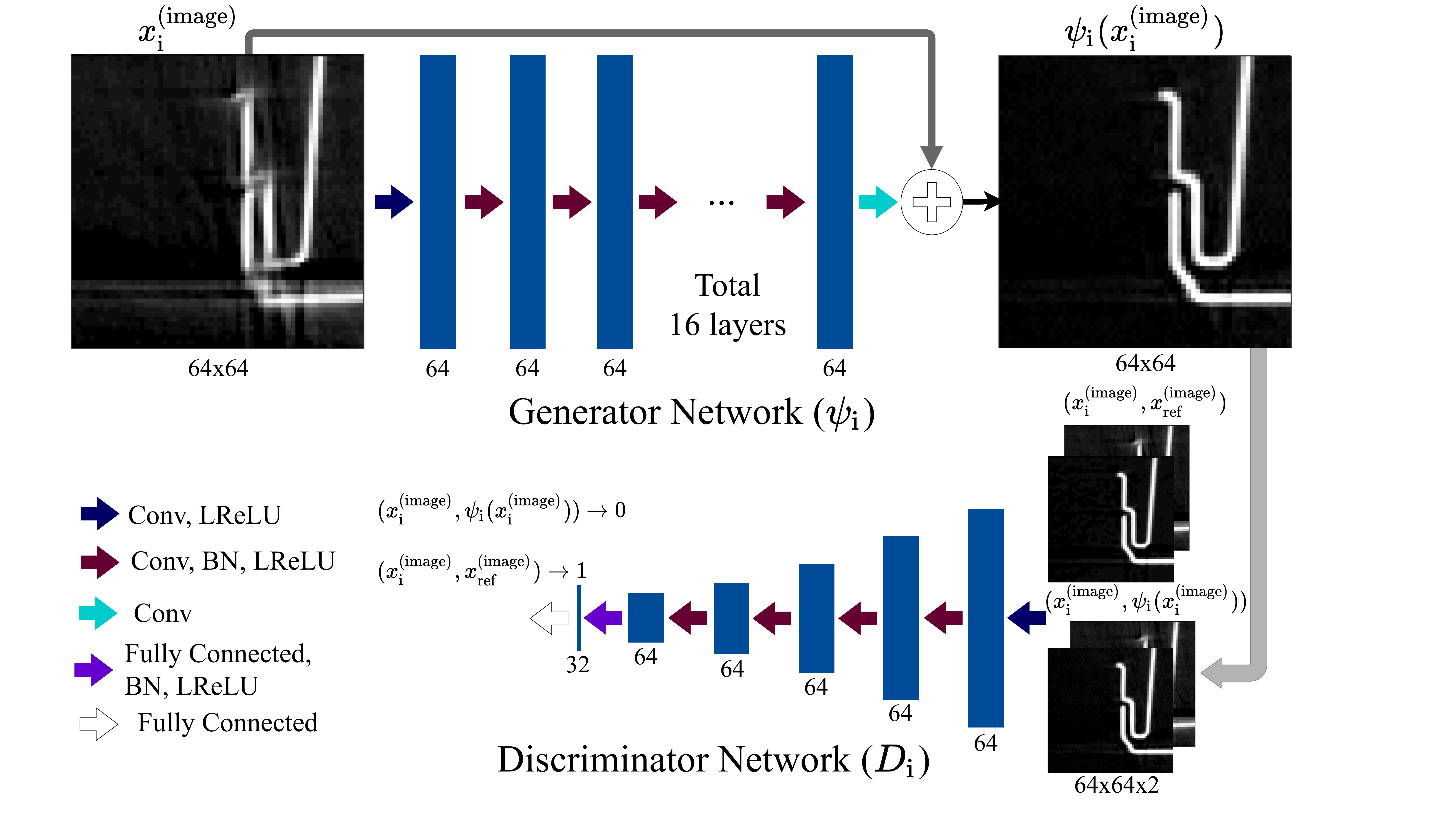}
	\caption{Overall architecture of the image-domain cGAN $\psi_\imageagent(\estimage_\imageagent)$. It learns patch-based image priors from a large security dataset. Here, $\estimage_\textnormal{ref}$ is a patch of a reference high-quality training image. The abbreviated terms in the Figure are defined here; Conv: 2D convolution, LReLU: leaky rectified linear unit, and BN: batch-normalization. }
	\label{fig:cgan_res_vdsr}
\end{figure*}

\subsection{Data-Domain cGAN $\psi_{\dataagent}(\estdata_\dataagent)$} \label{data_DL_model}

The implicit data model (\ref{eq:data-implicit}) uses a DNN $\psi_{\dataagent}(\estdata_\dataagent)$ to repeatedly enhance the current estimate of the data variable. This network uses a conditional generative adversarial network (cGAN) \cite{pix2pix17} for its structure and is based on the same network architecture used in \cite{ghani2019integrating} for data completion.  The data enhancement network $\psi_{\dataagent}$, however, is trained to learn a mapping from imperfect data estimates to target reference data, that is, to perform enhancement of the entire set of data. 

The network architecture of $\psi_{\dataagent}(\estdata_\dataagent)$ is presented in Figure~\ref{fig:cgan_unet}. The architecture consists of a Generator network and a Discriminator network. Both networks are trained using a combination of adversarial \cite{goodfellow2014generative} and mean squared error (MSE) loss. The Generator network follows the U-Net \cite{ronneberger2015u} architecture with $6$ down-sampling and $6$ up-sampling layers. We use $2$-pixel strided convolutions for down-sampling and transposed convolutions for up-sampling. All layers use $7\times7$ convolutional kernels. The generator $\psi_\dataagent$ has a theoretical effective receptive field (ERF) of $1135\times1135$ pixels. Such a large ERF is needed due to the non-local structure of the sensor data in the imaging applications of interest.

\subsection{Image-Domain cGAN $\psi_\imageagent(\estimage_\imageagent)$} \label{image_DL_model}

A cGAN is also used the image-domain prior model $\psi_\imageagent(\estimage_\imageagent)$. The architecture of this image-domain cGAN is given in Figure~\ref{fig:cgan_res_vdsr}. The Generator network architecture is inspired from \cite{kim2016accurate,zhang2017beyond}. It learns to estimate residual error by using a skip connection between the input and output of the last layer. The Generator and Discriminator networks are trained jointly using a combination of adversarial loss \cite{goodfellow2014generative} and MSE loss applied to image patches. The Generator network learns a mapping from lower-quality reconstructions to reference reconstructions. It uses a fully convolutional architecture, with $5\times5$ kernels, and a $1$-pixel strided convolutions. The Generator network architecture results in an ERF of $65\times65$ pixels.

\section{Experiments}
\label{sec:exp}

In this section we present experimental results of using our framework and provide comparisons to common alternatives that demonstrate the value of combining both data and image domain models. We focus on two canonical incomplete-data applications utilizing the incomplete data sensor agent from Section~\ref{sec:incompletedatamodel}: i) limited-angle CT, and ii) accelerated MRI. In both cases we use $4$ CE iterations of our DIPIIR method, $20$ CG iterations in solving (\ref{eq:missingdata_agent}), and Tensorflow in implementing the deep learning components. 

\subsection{Limited-Angle CT} \label{sec:LACT}

This section focuses on a $90^0$ limited-angle CT application and reports experimental results on a real checked-baggage dataset acquired using an Imatron C300 scanner \cite{crawford2014advances}. The field of view was $475\textnormal{mm} \times 475\textnormal{mm}$ with $130 keV$ peak source energy. The data was collected using a fan-beam geometry and was then rebinned as parallel beam observations with $720$ projection angles and $1024$ detector channels. Incomplete data were created by using projections in the limited range $[0^0, 90^0]$. Slices from $168$ bags were used for training and $21$ bags for testing. Slices with metallic objects were not considered for this work. The same data and training strategy was used for the data and image domain cGAN models as in \cite{ghani2019integrating}. The ASTRA toolbox \cite{van2016fast} was used for accelerated forward and back projection operations on a GPU.

For the explicit data-domain prior element $\proxvar_0^{\textnormal{(data)}}$ the output of a data completion DNN modeled on \cite{ghani2019integrating} was used. Details can be found in the supplementary material. The following additional parameters were used for the explicit data-domain prior model case: $\rho=0.5, \mu_\sensoragent=0.6, \mu_\imageagent=0.2, \mu_\dataagent = 0.2,\lambda_\sensoragent = 3.3\times 10^6$, and $\lambda_\dataagent=2$. For DIPIIR with the implicit data-domain prior model $\psi_{\dataagent}(\estdata_\dataagent)$, the model in Section~\ref{data_DL_model} was used with the following framework parameters: $\rho=0.35, \mu_\sensoragent=0.65, \mu_\imageagent=0.15, \mu_\dataagent = 0.20, \lambda_\sensoragent = 2.0\times 10^6$, and $\lambda_\dataagent=3.33$. 

The following initialization was used for Algorithm 1:
\begin{equation}
	x^{(0)} = 
	\left[
	\begin{array}{c}
	\text{FBP}(\proxvar_0^{\textnormal{(data)}})
	\\ 
	\hline
	\mathbf{A}_\text{unobs} \text{FBP}(\proxvar_0^{\textnormal{(data)}})
	\end{array}
	\right], 
	\hspace{.25in}
	\stackedcevar^{(0)}
	=
	\left[
	\begin{array}{c}
	x^{(0)} \\ x^{(0)} \\ x^{(0)}
	\end{array}
	\right]	
\end{equation}
where $\proxvar_0^{\textnormal{(data)}}$ is the previously discussed data completion estimate of the missing data described in the supplementary material and FBP denotes the conventional FBP image formation algorithm. While we use this initialization scheme for our experiments, our DIPIIR framework is not particularly dependent upon initialization in our experience.

The proposed DIPIIR framework is compared to six different image reconstruction approaches listed next:
\begin{description}[leftmargin=-1pt]
	\item[FBP:] The industry standard FBP algorithm applied to the original, incomplete data. We use the FBP implementation in the ASTRA toolbox \cite{van2016fast} with the Ram-Lak filter.
	
	\item[FBP+PP:] FBP combined with DNN-based post-processing (PP). This combination has been a popular theme in CT imaging research recently \cite{jin2017deep,yang2017low,li2017low,han2018framing}. For a fair comparison, the architecture and training strategy for the PP DNN network is the same as that as used for the image domain cGAN model, described in Section~\ref{image_DL_model}. 
	\item[DC+FBP:] Data completion pre-processing followed by FBP reconstruction has emerged as an alternative to PP \cite{lee2018deep,ghani2018deep,dong2019sinogram}. For a fair comparison, the architecture and training strategy of the data completion network is the same as that used for the explicit data-domain prior $\proxvar_0^{\textnormal{(data)}}$, as detailed in the supplemental material.  
	\item[DC+FBP+PP:] Data completion pre-processing, FBP inversion, and subsequent DNN post-processing. Such combination approaches have been shown to yield higher quality results as compared to using only pre-processing or only post-processing \cite{liang2018comparision}. For a fair comparison here, this approach combines the DNN-based data completion of the DC+FBP approach with the DNN-based post-processing of the FBP+PP approach. 
	\item[PnP-MBIR:] Plug-and-play MBIR described in \cite{venkatakrishnan2013plug,ye2018deep}. A model-based method that only includes image priors. The same image prior network has been used in the PnP-MBIR \cite{ye2018deep} and our DIPIIR method. The regularization parameter used for PnP-MBIR is: $\sigma^2=1.0 \times 10^{-7}$. 
	\item[DICE:] This is the method in \cite{ghani2019integrating} combining data and image models, but where the data estimate is not updated in a unified way. The parameters used for DICE are: $\rho=0.4, \mu_1=0.6, \mu_2=0.4$, and $\sigma^2=1.0\times10^{-8}$.
\end{description}
 
\begin{figure}[tb]
    \centering
    \includegraphics[height=0.7\textheight]{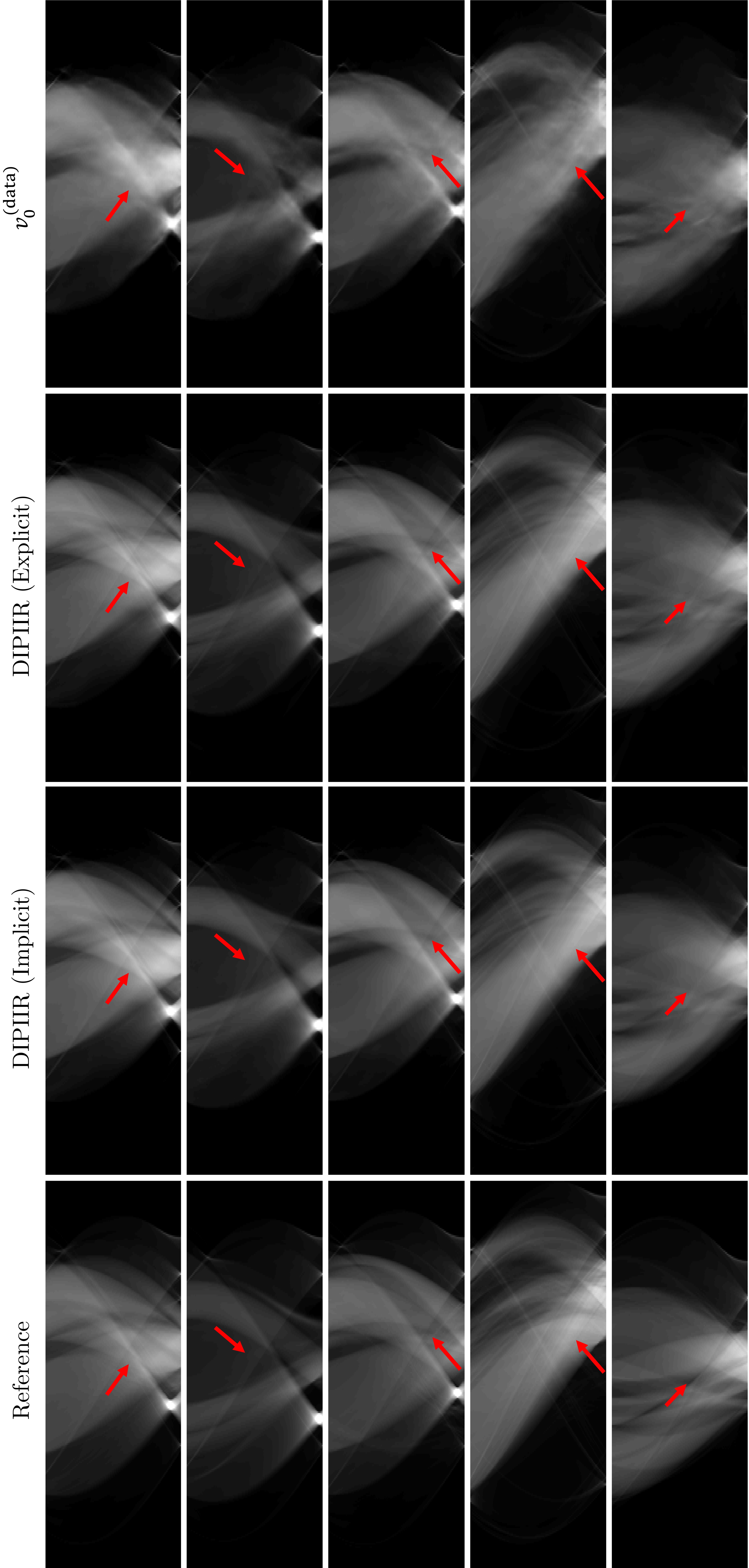}
    \caption{Data estimates $\estdata$ are presented. Each column is a different example. Only the unobserved part of the sinogram data are shown here.}
    \label{fig:sino}
\end{figure}

Estimates of the missing sinogram data $\estdata$ are presented in Figure~\ref{fig:sino} for the various methods that create them (not all methods generate such estimates). The result of the DC DNN estimate $\proxvar_0^{\textnormal{(data)}}$, the final estimates produced with both the explicit and implicit DIPIIR method, and the reference data are presented. Each coumn presents results for a different example. Red arrows are used to highlight regions where the proposed DIPIIR approach appears to significantly improve the data estimate over simple data completion. The data completion estimates $\proxvar_0^{\textnormal{(data)}}$ are not bad, and capture high-level features but they suffer in non-smooth regions. The DIPIIR approach integrates data and image prior models and this integration of information appears to improve the estimates of the data variable. 

\begin{figure*}
    \centering
    \includegraphics[height=0.95\textheight]{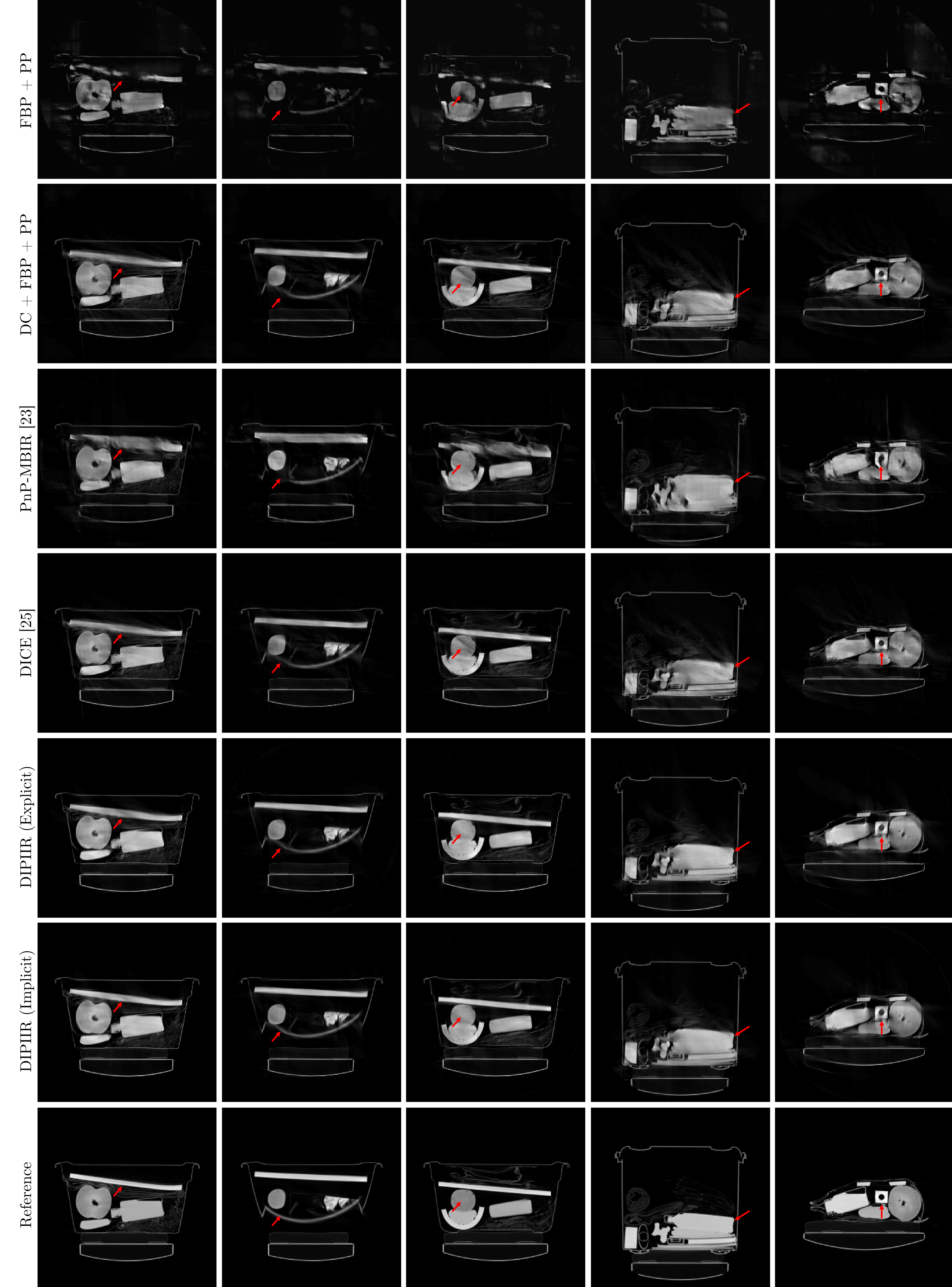}
    \caption{Image reconstruction results for a challenging $90^0$ limited angle CT problem are presented. The comparison methods are described in Section~\ref{sec:LACT}. Using only image-domain DL leaves residual artifacts and missing structures (first and third row). The proposed DIPIIR approach produces superior quality reconstructions, suppressing artifacts and successfully recover lost information.}
    \label{fig:recons}
\end{figure*}

Reconstruction results on several examples are presented in Figure~\ref{fig:recons}, where the various reconstruction methods are compared to the output of the DIPIIR framework. Red arrows again indicate areas where inclusion of both data and image priors lead to improvements. All of the approaches considered enhance image quality as compared to simple FBP reconstruction, however, many still leave perceptible residual artifacts and missing structural features. Methods using just an image-domain model appear to perform worse than methods which combine data and image domain models. The comparison to PnP-MBIR \cite{ye2018deep} is particularly interesting, since it uses a similar model-based framework and an image-domain learned prior model, however, it lacks the information derived from a data-domain model. The DIPIIR framework also appears to improve upon DICE \cite{ghani2019integrating}, showing the value of an balanced and integrated framework. A quantitative comparison of all the methods on the $484$ test examples is presented in Table~\ref{tab:rec} confirming the visual improvements provided by the DIPIIR method in Figure~\ref{fig:recons}. 

\begin{table}[tb]
  \centering
  \caption{CT reconstruction performance comparison.} 
    \begin{tabular}{|l|c|c|c|}
    \hline
 \textbf{Method} & \textbf{RMSE} (HU) & \textbf{PSNR} (dB) & \textbf{SSIM}\\ \hline
	FBP & $116$ & $22.49$ & $0.56$ \\ \hline
	FBP + PP   & $103$ & $23.32$ & $0.48$ \\ \hline
	DC + FBP   & $65$ & $27.53$ & $0.80$ \\ \hline
	DC + FBP +  PP & $60$ & $28.16$ & $0.76$ \\ \hline
	PnP-MBIR \cite{ye2018deep} & $78$ & $25.65$ & $0.79$ \\ \hline
	DICE  \cite{ghani2019integrating}  & $58$ &  $28.53$ &   $0.85$ \\ \hline
	DIPIIR (Explicit) & $54$ &  $29.03$ &   $0.86$ \\ \hline
	DIPIIR (Implicit)   & $\mathbf{51}$ &  $\mathbf{29.50}$ &   $\mathbf{0.87}$ \\ \hline
    \end{tabular}%
  \label{tab:rec}%
\end{table}%

\begin{figure*}
	\centering
	\includegraphics[width=0.9\textwidth]{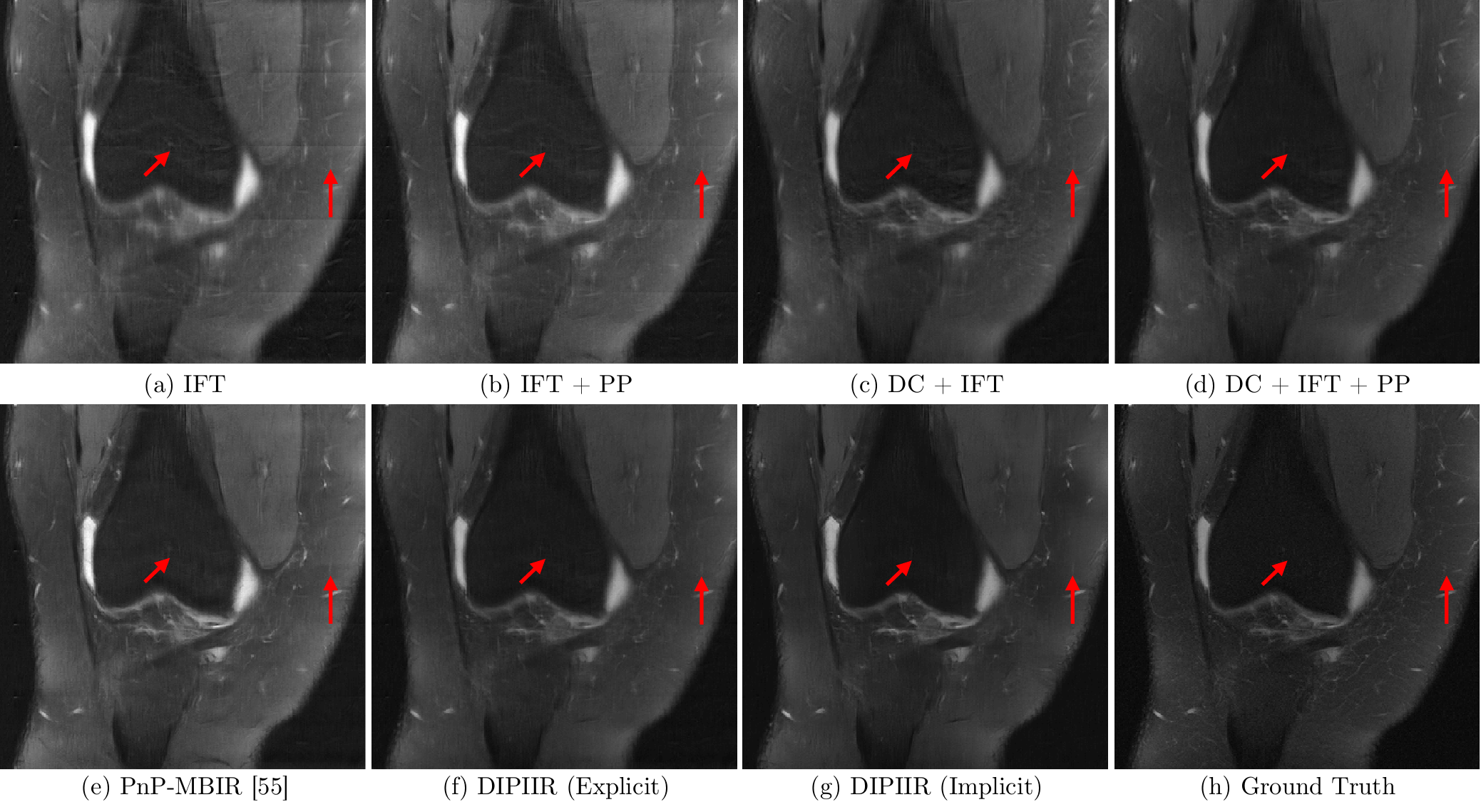}
	\vspace{-1em}
	\caption{MR image reconstruction results produced with different methods are presented. Red arrows highlight some of the issues in compared methods, where residual reverberating artifacts are visible, whereas our DIPIIR framework successfully suppress those artifacts and also improve image quality. Here, DC refers to data domain cGAN based Data Completion, PP refers to post-processing using image domain cGAN.}
	\label{fig:mri_rec_res}
\end{figure*}

\subsection{Accelerated MRI} \label{sec:AMRI}

In this section a $4\times$ accelerated MRI application is used as another incomplete data problem. Ground truth single channel knee MRI images from the fastMRI challenge \cite{zbontar2018fastmri} are used. Each image slice is $160 \textnormal{mm} \times 160 \textnormal{mm}$ with $0.5 \textnormal{mm} \times 0.5 \textnormal{mm}$ resolution. The 2D Fourier transform (FT) is used as a forward model to generate data followed by $4\times$ uniform sub-sampling in k-space with a $6\%$ auto-calibration signal (ACS) resulting in a net $3.64\times$ data reduction and acceleration. Image slices with almost no content were not used. From the training dataset split $973$ volumes were used, resulting in $28,904$ training image slices. The first $40$ volumes from validation split were used, resulting in $1151$ test slices. 

For data-domain (k-space) modeling, we follow the strategy of Han \etal \cite{han2019k} and break the complex k-space data into real and imaginary components. We also use their weighting strategy where data-domain DL models are trained on weighted k-space data. We use the same data-domain DNN architecture and learning scheme described in Section~\ref{data_DL_model} except for the following differences: \par
\begin{itemize}
    \item Two channel input and output k-space data (real and imaginary components) is used, resulting in input and output sizes of $320\times320\times2$. 
    \item The discriminator network ($D_\dataagent$) consists of $5$ convolutional layers and $1$ fully-connected layer.
    \item A pseudo-Huber loss function is used for pixel-loss with:  $L_2(e) = 4 \left( \sqrt{1+(e/2)^2} - 1\right)$, where $e$ is the pixel error.
    \item The Adam optimizer \cite{kingma2014adam} is used with learning rate $0.002$, batch size $32$, and trained for $100$ epochs.
\end{itemize}

A patch-based prior image model is learned using the same network architecture and learning scheme described in Section~\ref{image_DL_model}. Network inputs are cropped from images generated by zero-filling the under-sampled k-space data and applying the 2D inverse Fourier transform (IFT). Full-data reference images from the knee MRI dataset are used as ground truth. The explicit DIPIIR data prior variable $\proxvar_0^{\textnormal{(data)}}$ is based on a data completion estimate computed using a k-space DNN, as described in the supplementary material. For DIPIIR with an explicit data-domain prior model the following parameters are used: $\rho=0.45, \mu_\sensoragent=0.45, \mu_\imageagent=0.35, \mu_\dataagent = 0.20,\lambda_\sensoragent = 2.0\times 10^5$, and $\lambda_\dataagent=1$. For DIPIIR with an implicit data-domain prior model the parameters are set as follows:  $\rho=0.45, \mu_\sensoragent=0.45, \mu_\imageagent=0.35, \mu_\dataagent = 0.20, \lambda_\sensoragent = 2.0\times 10^5$, and $\lambda_\dataagent=1$. 

Similar to the CT example, the following initialization was used for Algorithm 1:
\begin{equation}
	x^{(0)} = 
	\left[
	\begin{array}{c}
	\text{IFT}(\proxvar_0^{\textnormal{(data)}})
	\\ 
	\hline
	\mathbf{A}_\text{unobs} \text{IFT}(\proxvar_0^{\textnormal{(data)}})
	\end{array}
	\right], 
	\hspace{.25in}
	\stackedcevar^{(0)}
	=
	\left[
	\begin{array}{c}
	x^{(0)} \\ x^{(0)} \\ x^{(0)}
	\end{array}
	\right]	
\end{equation}
where $\proxvar_0^{\textnormal{(data)}}$ is the previously specified data completion estimate of the missing data described in the supplementary material and IFT denotes the conventional inverse Fourier transform image formation operator. While we use this initialization scheme for our experiments, our DIPIIR framework is not particularly dependent upon initialization in our experience.

The proposed DIPIIR framework is compared to five different image reconstruction approaches as follows:
\begin{description}[leftmargin=-1pt]
	\item[IFT:] The standard inverse Fourier transform applied to the incomplete Fourier data with missing data filled by zeros. It is a common strategy used in MR. 
	\item[IFT+PP:] IFT of the zero-filled data, followed by DNN-based post-processing, following the strategy in \cite{lee2018deepMRI}. This represents a fast post-processing approach similar to what has been done in CT. For a fair comparison, the architecture and training strategy for the PP DNN network is the same as that used for the image domain cGAN model described in Section~\ref{image_DL_model} and used as the DIPIIR image domain prior network. 
	\item[DC+IFT:] Fourier data completion pre-processing followed by IFT reconstruction. This approach has been shown to produce high quality images in certain cases \cite{han2019k}. For a fair comparison, the architecture and training strategy of the data completion network is the same as that used for the explicit data-domain prior $\proxvar_0^{\textnormal{(data)}}$, as detailed in the supplemental material.
	\item[DC+IFT+PP:] Data completion pre-processing, IFT inversion, and subsequent DNN post-processing. This combination approach has produced higher-quality images and is popular in the MR literature \cite{eo2018kiki}. For a fair comparison here, this approach combines the DNN-based data completion of the DC+FBP approach with the DNN-based post-processing of the FBP+PP approach.
	\item[PnP-MBIR:] Plug-and-play MBIR method described in \cite{venkatakrishnan2013plug}. PnP-MBIR has been used for MR imaging applications \cite{ahmad2020plug} and produced high-quality results. It is a model-based method that only include image priors. The same image prior network has been used in the PnP-MBIR \cite{ahmad2020plug} and our DIPIIR method. The regularization parameter used for PnP-MBIR is: $\sigma^2 = 5 \times 10^{-6}$.
\end{description}

Images of the Fourier data estimates are not presented since it is difficult to draw any conclusions from qualitative images of that complex data. Qualitative reconstruction results from the various methods are presented on a test example in Figure~\ref{fig:mri_rec_res}. Severe ghosting artifacts are visible in the images produced by zero-filling and IFT alone. All methods considered here attempt to suppress these artifacts and improve image quality. Residual artifacts are visible in the images produced by all methods except the proposed DIPIIR approach, which not only successfully suppressed artifacts but also appears to improve overall image quality. The comparison to PnP-MBIR is especially interesting because it also exploits a physical sensing model, but only combines that with image-domain prior information. By integrating those models with data-domain information the proposed DIPIIR approach can improve image quality. Quantitative results obtained over the entire dataset are presented in Table~\ref{tab:mri_rec_res} and confirm the improvements and the potential of integrating complementary priors. 

\begin{table}[tb]
  \centering
  \caption{MR reconstruction performance comparison.}
    \begin{tabular}{|l|c|c|c|}
    \hline
 \textbf{Method} & \textbf{NMSE} & \textbf{PSNR} & \textbf{SSIM}\\ \hline
	IFT & $5.61\times10^{-2}$ & $27.72$ & $0.796$ \\ \hline
	IFT + PP   & $4.26\times10^{-2}$ & $28.65$ & $0.813$ \\ \hline
	DC + IFT   & $4.32\times10^{-2}$ & $28.90$ & $0.812$ \\ \hline
	DC + IFT +  PP & $3.03\times10^{-2}$ & $29.85$ & $0.823$ \\ \hline
	PnP-MBIR \cite{ahmad2020plug}& $2.51\times10^{-2}$ & $30.10$ & $0.822$ \\ \hline
	DIPIIR (Explicit) & $2.47\times10^{-2}$ &  $30.41$ &   $0.827$ \\ \hline
	DIPIIR (Implicit)   & $\mathbf{2.30\times10^{-2}}$ &  $\mathbf{30.57}$ &   $\mathbf{0.828}$ \\ \hline
    \end{tabular}%
  \label{tab:mri_rec_res}%
\end{table}%

\section{Other Applications}

The proposed DIPIIR framework is flexible and can be applied to a wide range of computational imaging applications. In this section we suggest how the sensor agent can be crafted, and in particular, choices of $\measdata$ and $\Amat$, for two additional classes of problems to illustrate how this might be accomplished. These problems are the subject of future work, so we merely show how the framework might accommodate them here. 

\subsection{Application to Deblurring}

A canonical problem is image deblurring from noisy data \cite{hansen2006deblurring}. These inversion problem are made challenging by the presence of noise in the observed data, so the raw data is often pre-processed to perform denoising which is then followed by subsequent inversion. Our framework provides a means to jointly do these tasks of denoising the data and estimating the underlying image. To that end, we let $\estdata$ represent noise-free (or noise-reduced) pseudo-data and then define the sensing vector $\measdata$ and sensing operator $\Amat$ as follows:

\begin{eqnarray}
\measdata 
& = &  
\left[
\begin{array}{c}
y_\text{noisy} \\ \hline
\underline{0}
\end{array}
\right] \\
\Amat 
&=&  
\left[ 
{\begin{array}{cc}
	\mathbf{A}_\text{blur}   & \mathbf{0}_p \\
	\mathbf{A}_\text{blur}   & -\mathbf{I}_p \\
	\end{array} } 
\right]
\end{eqnarray}
where $y_\text{noisy}$ is the observed, noisy data, $\mathbf{A}_\text{blur}$ captures the physical blurring from the underlying image $\estimage$ to the observed data. Incorporating these choices into (\ref{eq:model_agent}) and rewriting the resulting sensor agent with $W=I$ we obtain: 
\begin{align}
	\label{eq:noisydata_agent}
	&
	F_\sensoragent(\cevar_\sensoragent) = \\ \nonumber
	&
	\argmin_{\proxvar \geqslant 0} 	
	\| y_\text{noisy} - \mathbf{A}_\text{blur} \proxvar^\text{(image)} \|^2 
	+
	\| \proxvar^\text{(data)} - \mathbf{A}_\text{blur} \proxvar^\text{(image)} \|^2 
	\\ \nonumber
	& ~~~~~~~~~~
	+ 
	\lambda_\sensoragent \|\proxvar - \cevar_\sensoragent\|_2^2
\end{align}
The first term in (\ref{eq:noisydata_agent}) links the underlying image to the noisy data, while the second term couples the image estimate with the clean pseudo-data estimate through the observation model. 

Note that the same set up can be used for other problems with complete, but noisy data, by just changing the sensing operator. Examples would include low-dose CT \cite{chang2017modeling, ghani2018cnn}, MR spectroscopic imaging \cite{nguyen2012denoising}, MRI artifact correction \cite{jin2017mri}, and low-dose positron emission tomography (PET) imaging \cite{gong2018iterative}. 

\subsection{Application to Super-resolution}

Another interesting application is image super-resolution. In these problems a common model is that the observations are related to an underlying high-resolution image through the action of two operators -- a convolutional and shift-invariant blurring operator followed by a sub-sampling operator. One approach in this case would be to let $\estdata$ represent the blurred but unsampled, noise-free image and then define the sensing vector $\measdata$ and sensing operator $\Amat$ as follows:
\begin{eqnarray}
\measdata 
& = &  
\left[
\begin{array}{c}
y_\text{lowres} \\ \hline
\underline{0}
\end{array}
\right] \\
\Amat 
&=&  
\left[ 
{\begin{array}{cc}
	\mathbf{A}_\text{sub}\mathbf{A}_\text{blur}   & \mathbf{0}_p \\
	\mathbf{A}_\text{blur}   & -\mathbf{I}_p \\
	\end{array} } 
\right]
\end{eqnarray}
where $y_\text{lowres}$ are the low resolution,blurred and subsampled observations, $\mathbf{A}_\text{blur}$ is a shift-invariant convolutional blurring operator and $\mathbf{A}_\text{sub}$ is a subsampling operator. The resulting sensor agent for this case would then become:
\begin{align}
\label{eq:superres_agent}
&
F_\sensoragent(\cevar_\sensoragent) = \\ \nonumber
&
\argmin_{\proxvar \geqslant 0} 	
\| y_\text{lowres} - \mathbf{A}_\text{sub}\mathbf{A}_\text{blur} \proxvar^\text{(image)} \|^2 
\\ \nonumber
& ~~~~~~~~~~
+
\| \proxvar^\text{(data)} - \mathbf{A}_\text{blur} \proxvar^\text{(image)} \|^2 
\\ \nonumber
& ~~~~~~~~~~
+ 
\lambda_\sensoragent \|\proxvar - \cevar_\sensoragent\|_2^2
\end{align}
The first term in (\ref{eq:superres_agent}) connects the high-resolution image to the observations, while the second term connects the estimate of the underlying blurred, but unsampled image to the final high-resolution image. 

\section{Conclusion}

In this work, a novel framework for integration of data and image domain priors for image reconstruction is proposed. The consensus equilibrium framework is used to achieve this through state augmentation and the definition of three agents: a sensor agent related to the physical sensor model, an image-domain prior model agent and a data-domain prior model agent. Two canonical incomplete data applications were presented: limited angle CT, and accelerated MRI. Experimental results were provided on a real CT security dataset and a simulated accelerated MRI dataset. In these applications the inclusion of data-domain priors produced superior quality results and demonstrated the potential of the approach. The overall framework is quite general and can be applied to a wide range of computational imaging problems. Note that in the present paper we have chosen to incorporate explicit coupling between the image variable $\estimage$ and the data variable $\estdata$ into the sensor agent $F_\sensoragent$ through our choice of $\Amat$. While this choice makes the roles of the agents $F_\sensoragent$, $F_\dataagent$, and $F_\imageagent$ easier to understand, it would be straight forward to incorporate such coupling into $F_\dataagent$ and $F_\imageagent$ as well. Such possibilities are the focus of future work.

\small
\bibliographystyle{IEEEtran}
\bibliography{literature.bib}

\vfill 
\end{document}



\title{Data and Image Prior Integration for Image Reconstruction Using Consensus Equilibrium -- Supplementary Material}

\author{Muhammad Usman Ghani, ~\IEEEmembership{Student Member,~IEEE,}
        W. Clem Karl,~\IEEEmembership{Fellow,~IEEE} 
        \thanks{This material is based upon work supported by the U.S. Department of Homeland Security, Science and Technology Directorate, Office of University Programs, under Grant Award 2013-ST-061-ED0001. The views and conclusions contained in this document are those of the authors and should not be interpreted as necessarily representing the official policies, either expressed or implied, of the U.S. Department of Homeland Security.}
\thanks{M. U. Ghani and W. C. Karl are with the Department
of Electrical and Computer Engineering, Boston University, Boston,
MA, 02215 USA (e-mail: \texttt{\{mughani, wckarl\}@bu.edu)}.}}

\markboth{Submitted to IEEE Transactions on Computational Imaging}%
{Ghani and Karl: Data and Image Prior Integration for Image Reconstruction Using Consensus Equilibrium -- Supplementary Material}

\maketitle
This supplement provides additional results and information in support of our paper \cite{ghani2020dip}. Limited angle data completion and k-space data completion models used in the explicit data-domain prior models as well as in algorithm initialization are presented in Section~\ref{supp: lim_dc} and Section~\ref{supp: k_dc} respectively.

\section{Data Completion Deep Network} \label{supp:dc_net}
A conditional generative adversarial network (cGAN) \cite{pix2pix17} $\psi_{\textnormal{DC}}$ is used for data-domain completion in the generation of $\proxvar_0^{\textnormal{(data)}}$. The network architecture is taken from \cite{ghani2019integrating}. We use the same core network architecture to train data completion models for both limited angle CT and accelerated MRI application with some key differences discussed in relevant subsections. The data completion network is trained to learn a mapping from incomplete sensor data to reference complete data.\par

Our data-completion cGAN consists of a Generator network and a Discriminator network. We train them both using a combination of adversarial \cite{goodfellow2014generative} and traditional pixel loss. The Generator network follows the U-Net \cite{ronneberger2015u} architecture with $6$ down-sampling and $6$ up-sampling layers. We use $2$-pixel strided convolutions for down-sampling and transposed convolutions for up-sampling. All layers use $7\times7$ convolutional kernels. The generator $\psi_{\textnormal{DC}}$ has a theoretical effective receptive field (ERF) of $1135\times1135$ pixels. A large ERF is needed due to non-local structure of the sensor data under consideration. The same training strategy as in \cite{ghani2019integrating} is used.\par

\subsection{Limited Angle CT} \label{supp: lim_dc}

\begin{figure*}
	\centering
	\includegraphics[width=0.9\textwidth]{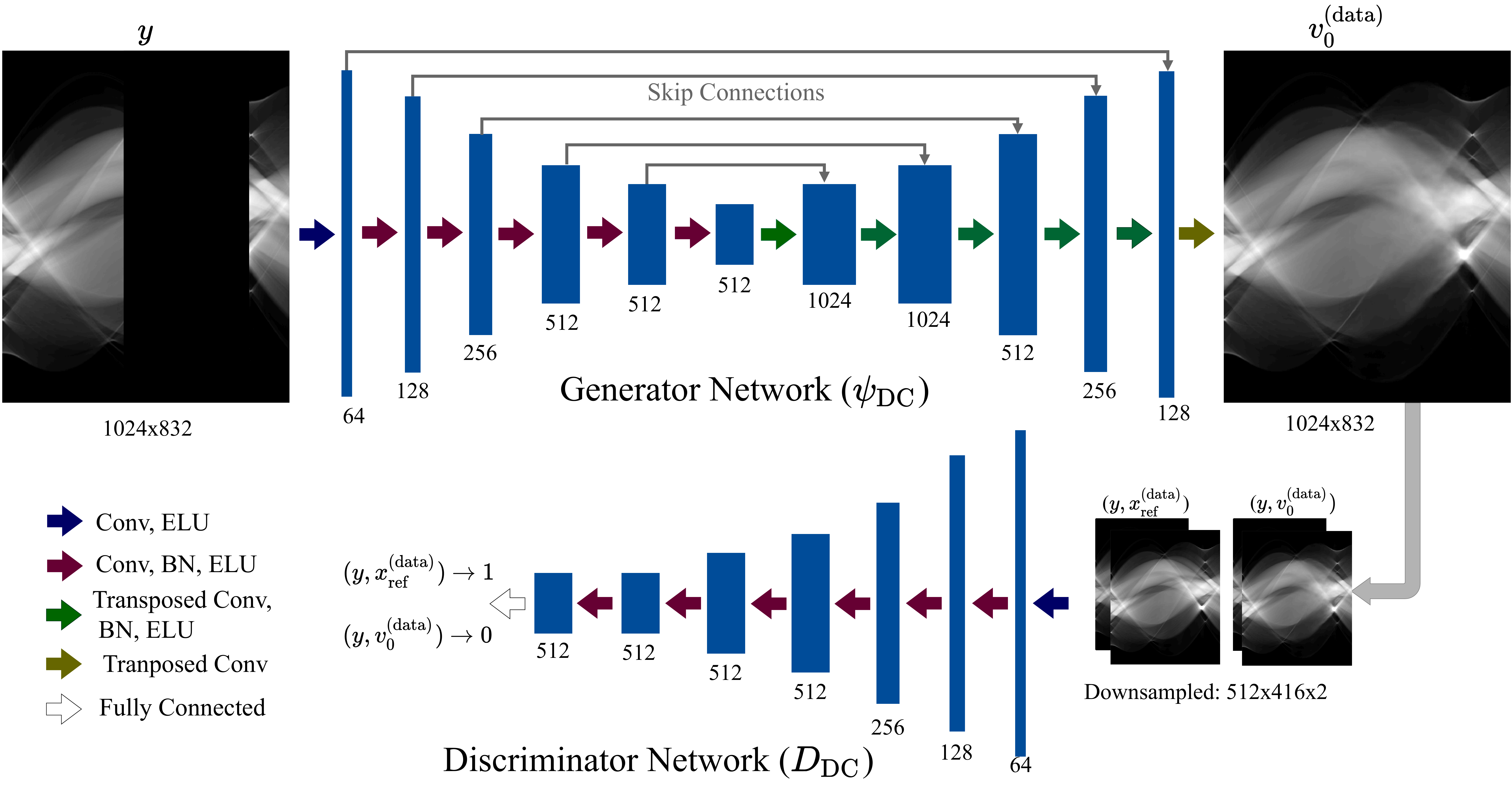}
	\caption{Overall architecture of the data completion cGAN is presented. Here, $\estdata_\textnormal{ref}$ is the reference high-quality sensor data. The abbreviated legends in the Figure are defined here; Conv: 2D convolution, ELU: exponential linear unit, BN: batch-normalization, and 2D Transposed Conv: transposed convolution.}
	\label{fig:cgan_unet_dc}
\end{figure*}
The network architecture of the data-domain cGAN is presented in Figure~\ref{fig:cgan_unet_dc}. We use the mean square error (MSE) pixel loss. In \cite{ghani2019integrating}, zero padding was used on original $1024\times720$ sinogram data to fit the network input and output dimensions. Here, the periodic nature of sinogram data is used instead of zero padding. In creating the completion DNN the input and reference projection data was extended beyond using a periodic boundary condition appropriate to sinograms. This boundary treatment was seen to improve the data-domain estimates compared to simple zero padding. 

\subsection{Accelerated MRI} \label{supp: k_dc}

\begin{figure*}
	\centering
	\includegraphics[width=0.9\textwidth]{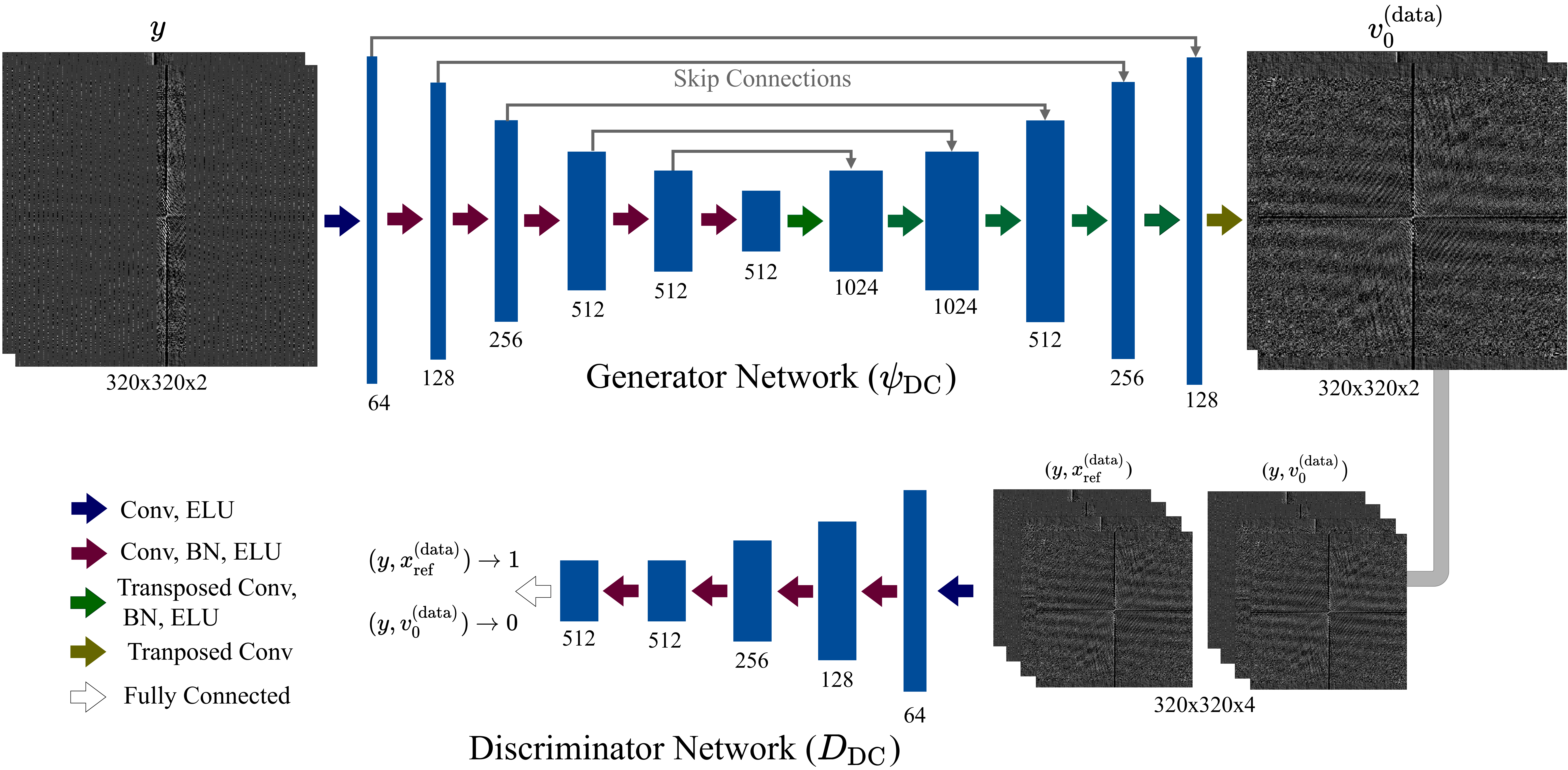}
	\caption{Overall architecture of our k-space data completion cGAN is presented. Here, $\estdata_\textnormal{ref}$ is the reference high-quality k-space data. The abbreviated legends in the Figure are defined here; Conv: 2D convolution, ELU: exponential linear unit, BN: batch-normalization, and 2D Transposed Conv: transposed convolution.}
	\label{fig:cgan_unet_dc_mri}
\end{figure*}

The k-space data completion CGAN used in this work is presented in Figure~\ref{fig:cgan_unet_dc_mri}. The strategy of Han \etal \cite{han2019k} is used, where complex k-space data is split into 2-channel real and imaginary data. Further, the data weighting scheme in Han \etal \cite{han2019k} to weight input and target data is used. The same training scheme as for the limited CT data completion problems is used except for the following differences: \par
\begin{itemize}
    \item Two channel input and output k-space data (real and imaginary data) are used resulting in input and output sizes of $320\times320\times2$. 
    \item The discriminator network ($D_\dataagent$) consists of $5$ convolutional layers and $1$ fully-connected layer.
    \item A pseudo-Huber loss function for pixel-loss is used with:  $L_\delta(e) = \delta^2 \left( \sqrt{1+(e/\delta)^2} - 1\right)$, where $e$ is the pixel error, and we use $\delta=2$.
    \item The Adam optimizer \cite{kingma2014adam} is used with learning rate $0.002$ with batch size $32$, and training for $100$ epochs.
\end{itemize}

\small
\bibliographystyle{IEEEtran}
\bibliography{literature.bib}